\documentclass[aps,prb,twocolumn,superscriptaddress,amsmath,amssymb]{revtex4}

\usepackage{graphicx}

\begin{document}


\title{Superconducting properties of Tl-doped PbTe}

\author{Y. Matsushita}
\author{P. A. Wianecki}
\affiliation{Department of Materials Science and Engineering and
Geballe Laboratory for Advanced Materials,\\
Stanford University, Stanford, California 94305}
\author{T. H. Geballe}
\author{I. R. Fisher}
\affiliation{Department of Applied Physics and Geballe Laboratory
for Advanced Materials,\\
Stanford University, Stanford, California 94305}

\date{\today}

\begin{abstract}
Tl-doped PbTe (Pb$_{1-x}$Tl$_{x}$Te) is an anomalous superconductor
with a remarkably high maximum $T_c$ value given its relatively low
carrier concentration.  Here, we present results of systematic
measurements of superconducting parameters for this material, for Tl
concentrations up to $x=1.4\%$.  We find that it is a Type II,
weak-coupled BCS superconductor in the dirty limit and discuss
implications for the applicability of the charge Kondo model
recently proposed to account for superconductivity in this system.
\end{abstract}


\maketitle

\section{Introduction}

Tl-doped PbTe (Pb$_{1-x}$Tl$_{x}$Te) is a degenerate semiconductor
with a small carrier concentration of $\sim10^{20}$ holes/cm$^{3}$
or less. However, for Tl concentrations $x$ beyond a critical value
$x_c \sim 0.3 \%$ it is observed to
superconduct,\cite{Matsushita_2005} with a maximum $T_c$ of 1.5 K
for the highest Tl concentrations,\cite{Nemov_1998} comparable to
more metallic systems. Furthermore, thallium is the only impurity
known to cause superconductivity in PbTe, even though other
impurities are able to dope to similar carrier concentrations and
similar densities of states. Given the anomalously high maximum
$T_c$ value, combined with the unusual concentration dependence,
there has been considerable discussion as to the role that the Tl
impurities play in the superconductivity of this
material.\cite{Moizhes_1983, Schuttler_1989, Hirsch_1985,
Krasinkova_1991, Dzero_2005}

PbTe has a rocksalt structure and has been treated with reasonable
success using ionic models (i.e.,
Pb$^{2+}$Te$^{2-}$).\cite{Weiser_1981} Thallium impurities
substitute on the Pb site, and calculations have shown that Tl$^{+}$
is more stable than Tl$^{3+}$ in the PbTe lattice.\cite{Weiser_1981}
This implies that Tl impurities will act as acceptors, and indeed
Hall measurements confirm that for small doping levels the hole
concentration increases by one hole for every Tl
impurity.\cite{Kaidanov_1985, Dzero_2005}  Carrier freeze-out is not
observed to the lowest temperatures, indicating that the dopant
atoms do not behave as hydrogen-like impurities due to the large
static dielectric constant of the host PbTe.\cite{Nimtz}  However,
for concentrations beyond a characteristic value the Hall number is
observed to rise at a much slower rate with $x$ and does not
increase beyond $\sim 10^{20}$ cm$^{-3}$,\cite{Matsushita_2005,
paper3} suggesting that the additional impurities act in a
self-compensating manner. Significantly, within the uncertainty of
these measurements, this characteristic concentration is the same as
$x_c$, the critical concentration required for superconductivity. It
is remarkable that as $x$ is increased beyond $x_c$, $T_c$ rises
linearly over two orders of magnitude from 15 mK for $x \sim0.3 \%$
to 1.5 K for $x \sim 1.5\%$, while the hole concentration varies by
less than a factor of two.

This behavior, combined with the absence of any detectable magnetic
impurities in the diamagnetic susceptibility, has been interpreted
as evidence that the Tl impurities are present in a mixed valence
state composed of a mixture of both Tl$^{+}$ and Tl$^{3+}$ valences
for $x > x_c$. We recently argued that anomalies in the normal state
resistivity of Tl-doped PbTe, that are present only for
superconducting samples ($x>x_c$) and not for nonsuperconducting
samples ($x<x_c$),\cite{Matsushita_2005, Fisher_2005} might be
associated with a charge Kondo effect involving these degenerate Tl
valence states. Within such a scenario, the quantum valence
fluctuations associated with the Tl impurities also provide a
possible pairing mechanism for holes in the valence band of the host
PbTe.\cite{Dzero_2005}

In light of the anomalous behavior of Tl-doped PbTe we have
investigated the superconducting properties of single crystal
samples for a range of Tl concentrations up to the solubility limit
of approximately 1.5$\%$. In this paper, we present measurements of
the heat capacity and $H_{c2}$ and the resulting estimates for
coherence length, penetration depth, Ginzburg-Landau parameter, and
critical fields. Our measurements show that the material is a Type
II, weak-coupled BCS superconductor in the dirty limit. We discuss
implications of these observations for the charge Kondo model.

\section{\label{sec:ii}Sample preparation and experimental methods}

Single crystals of Pb$_{1-x}$Tl$_x$Te were grown by an unseeded
physical vapor transport method. Polycrystalline source material was
synthesized by combining PbTe, Te, and either Tl$_2$Te or elemental
Tl in appropriate ratios and sintering at 600$^\circ$C, regrinding
between successive sintering steps. For the crystal growth, broken
pieces of source material were sealed in an evacuated quartz ampoule
and placed in a horizontal tube furnace held at 750$^\circ$C for
7--10 days. A small temperature gradient of approximately
1--2$^\circ$C/cm allowed crystals to nucleate and grow at one or
both of the cooler ends of the ampoule. Each vapor growth produced
several well-formed crystals up to a few millimeters in size that
could be cut and cleaved to prepare bars for thermodynamic and
transport measurements. The thallium content was measured by
Electron Microprobe Analysis (EMPA) using PbTe, Te, and Tl$_2$Te
standards. Errors in Tl content $x$ shown in subsequent figures
reflect the uncertainty of the microprobe method for such low dopant
concentrations.  The Tl concentration for individual samples was
observed to be homogeneous within the uncertainty of this
measurement.

The heat capacity of single crystal samples was measured using a
thermal relaxation technique in a Quantum Design Physical Property
Measurement System. Crystals with a mass of approximately 10--15 mg
were prepared with a flat surface for good thermal contact to the
sample platform. Measurements were made in zero applied field and in
a field $H > H_{c2}$ (typically $H$ = 0.5--1 T). The field was
oriented at an arbitrary angle to the crystal axes, depending on the
orientation of the flat surface.

The upper critical field $H_{c2}$ was measured for several values of
the Tl content $x$ by following resistive transitions as a function
of temperature for different applied magnetic fields. The
resistivity was measured using geometric bars cleaved from the
larger as-grown crystals, such that the current flowed along the
[100] direction while the magnetic field was oriented parallel to
the equivalent [001] direction.  Electrical contact was made using
Epotek H20E silver epoxy on sputtered or evaporated gold pads and
showed typical contact resistances of 1--4 $\Omega$. Resistivity
measurements were made at 16 Hz and with current densities in the
range of 25 mA/cm$^2$ (corresponding to a current of 100 $\mu$A for
low-temperature measurements) to 1 A/cm$^2$ at higher temperatures.
To check for heating effects, resistivity data were taken for
different current densities and for warming and cooling cycles for
each sample.

\section{\label{sec:iii}Results}

\begin{figure}[!]
\includegraphics{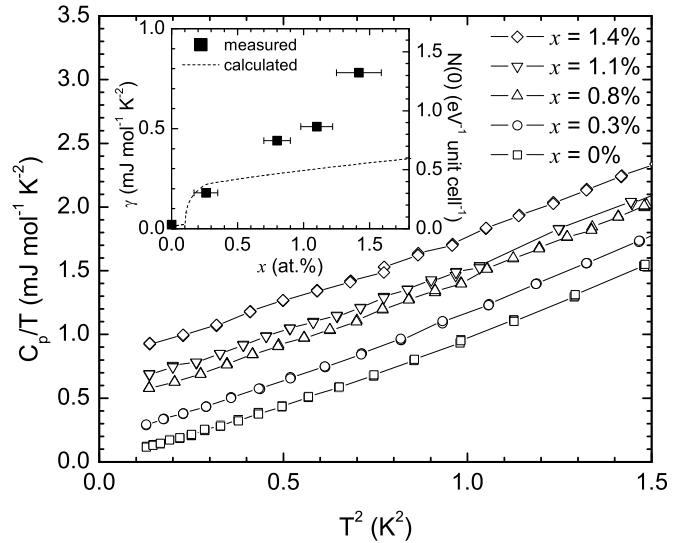}
\caption{\label{fig:Fig1} Heat capacity of Pb$_{1-x}$Tl$_x$Te single
crystals, shown as $C_p/T$ versus $T^2$, for representative Tl
concentrations.  For superconducting samples, data were taken in an
applied field $H > H_{c2}$.  Inset shows electronic contribution
$\gamma$ (left axis) and density of states at the Fermi level $N(0)$
(right axis) as a function of Tl concentration $x$. Dashed line
shows values calculated from known PbTe band parameters and measured
values of the Hall number, as described in the main text.}
\end{figure}

Heat capacity data for representative Tl concentrations are shown in
Fig.~\ref{fig:Fig1} as $C_p/T$ versus $T^2$ for applied fields that
totally suppress the superconductivity. For all samples there is
considerable curvature in the data even at low temperatures,
presumably due to the relatively low Debye temperature $\Theta_D$ of
PbTe.  Data were fit to $C/T = \gamma + \beta T^2 + \delta T^4$ from
the base temperature (0.3 K) up to 1 K. From
$\beta=N(12\pi^4/5)R\Theta^{-3}$, where $R=8.314$ J/(mol K) and
$N=2$ for PbTe, we estimate $\Theta_D = 168 \pm 4$ K for $x=0\%$,
which is consistent with previous
reports.\cite{Ravich,Chernik_1981a} Thallium substitution does not
substantially affect this value but causes a clear increase in the
electronic contribution $\gamma$, suggesting a rapid rise in the
density of states with $x$. Values of $\gamma$ obtained from the
above fits are shown in the inset to Figure 1 as a function of Tl
concentration $x$ and are in broad agreement with previously
published values for polycrystalline samples.\cite{Chernik_1981b}

\begin{figure}[t!]
\includegraphics{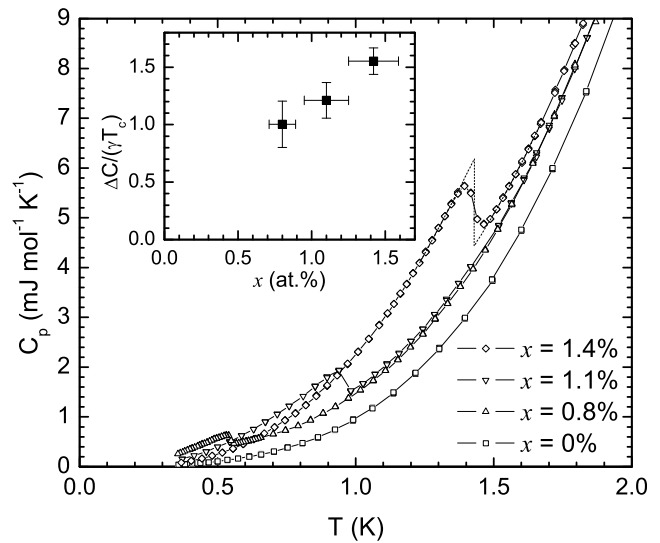}
\caption{\label{fig:Fig2}$C_p$ versus $T$ in zero applied field
showing the superconducting anomaly for several Tl concentrations
$x$. Dashed lines show the geometric construction used to obtain
$\Delta C$ and $T_c$ for $x=1.4\%$.  Inset shows $\Delta C$/$\gamma
T_c$ as a function of $x$.  Uncertainty in $\Delta C$/$\gamma T_c$
is derived principally from errors in the geometric construction
used to estimate $\Delta C$.}
\end{figure}

Heat capacity data in zero field are shown in Figure 2 for
representative Tl concentrations with $T_c$ above 0.3 K. $T_c$
values were obtained from the midpoint of the heat capacity anomaly
and agree well with data obtained from resistive transitions (Figure
3). The jump at $T_c$, $\Delta C$, can be estimated using a standard
geometric construction extrapolating normal state and
superconducting state behaviors towards $T_c$, as indicated by
dashed lines for $x$ = 1.4$\%$ in Figure 2. Resulting estimates of
$\Delta C / \gamma T_c$ are shown in the inset to Figure 2 as a
function of Tl concentration $x$. The value for the highest Tl
concentration, $x$ = 1.4$\%$, is $\Delta C/\gamma T_c$ = 1.55 $\pm$
0.12, which is close to the weak coupling BCS result of 1.43. As $x$
is reduced, the data show a small but significant systematic
variation, tending towards a smaller value for smaller Tl
concentrations. The smallest value, 1.00 $\pm$ 0.20, is recorded for
$x=0.8\%$, which is the lowest Tl concentration for which we can
confidently extract $\Delta C$ given the base temperature of our
instrument.

\begin{figure}
\includegraphics{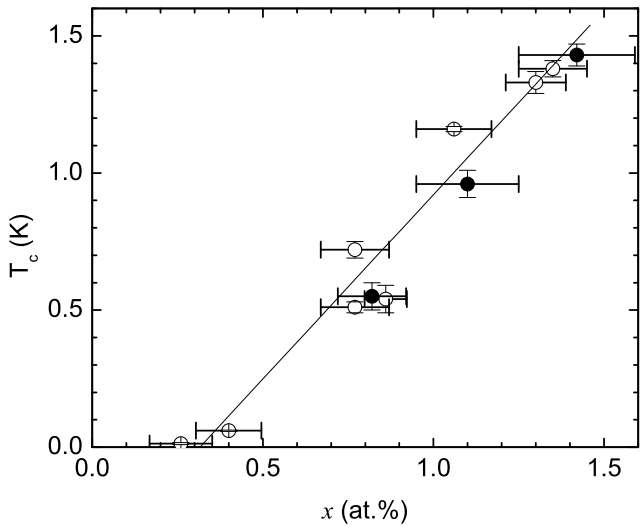}
\caption{\label{fig:Tcvx}Superconducting transition temperatures
$T_c$ as a function of Tl concentration $x$. Solid symbols are
obtained from heat capacity measurements, and open symbols are from
resistivity data.  Line is drawn to guide the eye.}
\end{figure}

\begin{figure}
\includegraphics{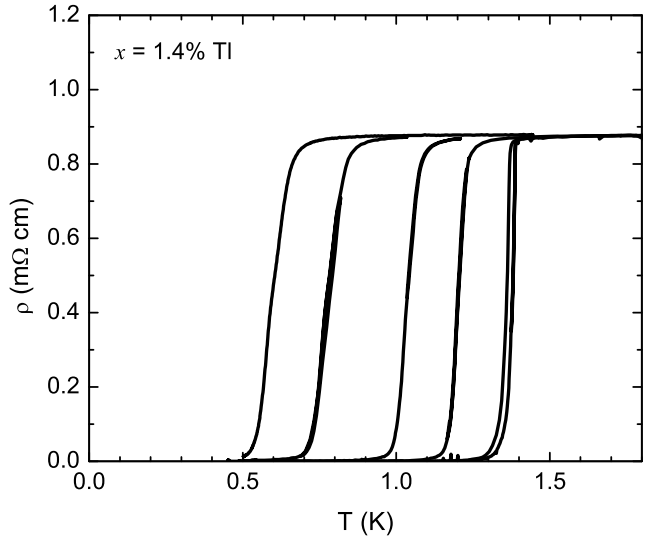}
\caption{\label{fig:trans_vs_field}Representative resistivity data
for $x=1.4\%$, showing the superconducting transition as a function
of temperature for different magnetic fields (0, 108, 1083, 2166,
3791, and 4875 Oe) applied parallel to the [001] direction.}
\end{figure}

\begin{figure}
\includegraphics{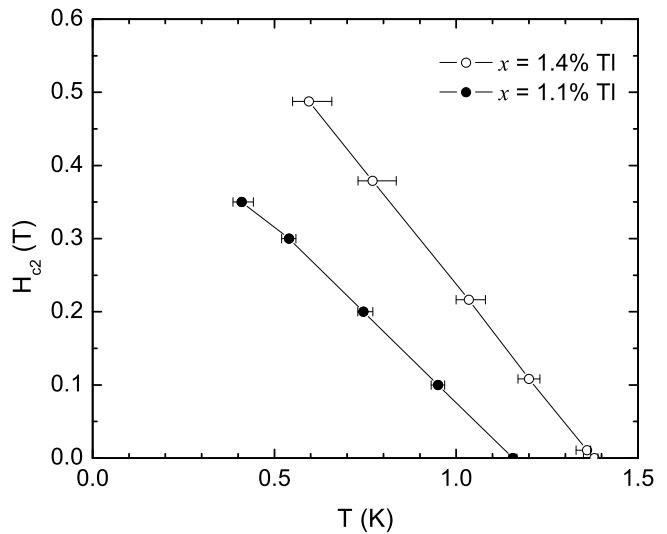}
\caption{\label{fig:hc2}Temperature dependence of $H_{c2}$ for
$x=1.1\%$ and $1.4\%$ with the field parallel to the [001]
direction. Error bars were determined as described in the main text.
Lines are drawn to guide the eye.}
\end{figure}

The upper critical field $H_{c2}(T)$ was determined from resistivity
measurements for several Tl concentrations. Representative data,
showing the uniform suppression of the superconducting transition in
an applied field, are shown in Fig.~\ref{fig:trans_vs_field} for $x$
= 1.4$\%$. An estimate of $T_c$ was obtained from the midpoint of
the resistive transition for each applied field. Resulting $H_{c2}$
curves are shown in Fig.~\ref{fig:hc2} for $x$ = 1.1$\%$ and
1.4$\%$. Error bars indicate the width of the superconducting
transition measured by the difference in temperature between $10\%$
and $90\%$ of the resistive transition. The upper critical field at
zero temperature $H_{c2}(T=0)$ can be estimated from these data
using the Werthamer-Helfand-Hohenberg approximation
\cite{Werthamer_1966}
\begin{equation}
H_{c2}(0)=0.69(dH_{c2}/dT)_{T_c}T_c.
\end{equation}
Resulting values for $x$ = 1.1$\%$ and 1.4$\%$ are approximately
3900 Oe and 6000 Oe respectively (Table~\ref{tab:table1}) and are
consistent with reasonable extrapolations of the lowest temperature
data in Fig. 5. The errors in $H_{c2}(0)$ listed in
Table~\ref{tab:table1} are estimated from the difference between the
above approximation and a parabolic fit to the observed data.

Superconducting parameters such as the coherence length and
penetration depth are dependent on the electron mean free path
$l=v_F\tau=v_F\mu m^*/e$, where $v_F$ is the Fermi velocity and
$\mu$ is the hole mobility.  From Hall effect measurements at 1.8 K,
we find that the Hall number $p_H$ is $\sim9 \times 10^{19}$
cm$^{-3}$ for $x = 1.1\%$ and $\sim1 \times 10^{20}$ for $x =
1.4\%$.\cite{paper3} Combining these data with measured values of
the residual resistivity,\cite{Matsushita_2005} we find the hole
mobility $\mu$ is approximately 100 cm$^2$V$^{-1}$s$^{-1}$ for $x =
1.1\%$ and 60 cm$^2$V$^{-1}$s$^{-1}$ for $x = 1.4\%$. Taking into
account the existence of both light and heavy holes at the $L$ and
$\Sigma$ points in the Brillouin zone
respectively,\footnote{\label{test}The valence band maximum is
centered at the $L$ points of the Brillouin zone and consists of
relatively light holes characterized by $m_l =0.31m_0$ and $m_t
=0.022m_0$ [R. Dornhaus, G. Nimtz, and B. Schlict, {\it Narrow-Gap
Semiconductors}, vol. 98 of {\it Springer Tracts in Modern Physics}
(Springer-Verlag, New York, 1983)]. Small deviations from
nonparabolicity can be safely ignored in estimating approximate
values for the Fermi level. Somewhat less is known of the secondary
band maximum centered at the $\Sigma$ points in the Brillouin zone.
Reasonable estimates were reported close to $m_{\Sigma} \sim m_0$
with an anisotropy of $\sim$10 [B.~F. Gruzinov, I.~A. Drabkin, and
Yu.~I. Ravich, Sov. Phys. Semicond. {\bf 13}, 315 (1979)].  These
holes are substantially more massive and therefore dominate the
density of states. In estimating the Fermi energy and other
electronic parameters from Hall effect measurements, we have assumed
that the band offset (170 meV) does not vary with $x$.} we assume
the elastic scattering limit holds at low temperatures such that
$l_L = l_{\Sigma}=l$.  The Fermi level then lies in the range
190--210 meV below the top of the valence band. Consequently,
average values of $v_F$ are $\sim 1.4\times 10^6$ m/s for the $L$
holes and $1 \times 10^5$ m/s for the $\Sigma$ holes. The resulting
values for $l$ are listed in Table \ref{tab:table1}. The principle
contribution to the uncertainty in this quantity arises from errors
in the geometric factor used to calculate resistivity of samples.
Propagation of this error is the dominant effect in the
uncertainties of subsequent derived quantities, including $\xi_0$
and $\lambda_{\mathrm{eff}}$.

The Ginzburg-Landau coherence length $\xi(0)$ is calculated from
$H_{c2}(0)$ by
\begin{equation}
H_{c2}(0)=\frac{\Phi_0}{2\pi\xi^2(0)},
\end{equation}
where $\Phi_0=2.0678\times10^{-15}$ T$\,$m$^2$.  Estimates for
$\xi(0)$ are 290 \AA{} for $x=1.1\%$ and 240 \AA{} for $x=1.4\%$
(Table~\ref{tab:table1}) and should be independent of orientation
for this cubic material. The small values of $l$ implies that the
material is in the dirty limit with $l < \xi_0$. Therefore, the
intrinsic coherence length $\xi_0$ can be extracted from the
approximation $\xi(0) \sim (l\xi_0)^{1/2}$, where the values are
listed in Table~\ref{tab:table1}. In comparison, the BCS expression
for $\xi_0$ is
\begin{equation}
\xi_0=\frac{\alpha\hbar v_F}{k_B T_c},
\end{equation}
where the BCS value of $\alpha$ is 0.18.  Using values of $\xi_0$
derived from the dirty limit approximation, we find $v_F$ estimated
from this formula (given in Table~\ref{tab:table1}) is between those
calculated separately for the $L$ and $\Sigma$ holes. This is
consistent with a mixed contribution from both carrier types due to
the substantial scattering implied from the short mean free path.

The London penetration depth for two carrier types is given by
\begin{equation}
\frac{1}{\lambda_L^2}=\frac{\mu_0 n_L e^2}{m_L}+\frac{\mu_0
n_{\Sigma} e^2}{m_\Sigma},
\end{equation}
where the superfluid densities $n_L$ and $n_\Sigma$ are approximated
as the carrier densities for each carrier type, and $m_L$ and
$m_\Sigma$ are the effective masses of each band.  The corresponding
values of $\lambda_L$ are listed in Table~\ref{tab:table1} and are
almost independent of orientation. In the dirty limit, we can
estimate the effective penetration depth from
\begin{equation}
\lambda_{\mathrm{eff}}=\lambda_L(\xi_0/l)^{1/2},
\end{equation}
values of which are given in Table~\ref{tab:table1}.  These
estimates are in good agreement with microwave conductivity
measurements that show $\lambda(0) \sim 3$ $\mu$m for
$x=1.4\%$.\cite{Ormeno} Finally, we find the Ginzburg-Landau
parameter using $\kappa=\lambda_{\mathrm{eff}}/\xi(0)$ and estimate
$H_c$ and $H_{c1}$ from the relationships $H_{c2}=\sqrt{2}\kappa
H_c$ and $H_{c1}=\frac{H_c}{\sqrt{2}\kappa}\ln\kappa$
(Table~\ref{tab:table1}).

\begin{table}
\caption{\label{tab:table1}Superconducting parameters of
Pb$_{1-x}$Tl$_x$Te for two representative Tl concentrations.}
\begin{ruledtabular}
\begin{tabular}{ccc}
&$x=1.1$ at.$\%$ &$x=1.4$ at.$\%$\\
\hline
$T_c$ &1.16 $\pm$ 0.01 K &1.38 $\pm$ 0.03 K\\
$H_{c2}(0)$ &0.39 $\pm$ 0.04 T &0.60 $\pm$ 0.07 T\\
$l$ &32 $\pm$ 8 \AA &19 $\pm$ 5 \AA\\
$\xi(0)$ &290 $\pm$ 15 \AA &240 $\pm$ 14 \AA\\
$\xi_0$ &2600 $\pm$ 700 \AA &3000 $\pm$ 850 \AA\\
$v_F$ &$2.2 \pm 0.6 \times10^5$ m/s &$3.0 \pm 0.8 \times10^5$ m/s\\
$\lambda_L$ &1600 $\pm$ 80 \AA &1500 $\pm$ 120 \AA\\
$\lambda_{\mathrm{eff}}$ &1.4 $\pm$ 0.4 $\mu$m &1.9 $\pm$ 0.5 $\mu$m\\
$\kappa$ &48 $\pm$ 12 &79 $\pm$ 20\\
$H_c$ &57 $\pm$ 14 Oe &54 $\pm$ 13 Oe\\
$H_{c1}$ &3 $\pm$ 0.8 Oe &2 $\pm$ 0.5 Oe\\
\end{tabular}
\end{ruledtabular}
\end{table}

\section{\label{sec:iv}Discussion}

The above results indicate that Tl-doped PbTe is a Type II, BCS
superconductor in the dirty limit, which is not too surprising given
that the material is a doped semiconductor. To a large extent this
observation rules out the possibility of more exotic scenarios for
the superconductivity, such as condensation of preformed pairs, at
least for the highest Tl concentrations. Here, we discuss some
implications for the charge Kondo model that we have previously
proposed for this material and consider alternative explanations.
First, we briefly reiterate the salient features of the charge Kondo
model relevant to understanding these data.

The idea of a charge Kondo effect associated with degenerate valence
(charge) states of a valence-skipping element was first discussed by
Taraphder and Coleman,\cite{Taraphder_1991} and was later
re-examined in the limit of dilute impurities for the case $T_c$
$\sim$ $T_K$ by Dzero and Schmalian.\cite{Dzero_2005} Weak
hybridization of these degenerate impurity states with conduction
electrons (or in the case of Tl-doped PbTe, with valence band
holes), results in a Kondo-like effect with various parallels to the
more common magnetic case. Here, the pseudospins correspond to zero
or double occupancy of an impurity orbital, which can be described
in terms of a negative effective $U$. The degeneracy of the two
valence states in PbTe is not accidental but emerges naturally from
the doping effect of the Tl impurities themselves.\cite{Dzero_2005}
For values of the chemical potential less than a characteristic
value $\mu^*$, the impurities are all present as one valence
(Tl$^+$), which act to dope the material. As more impurities are
added, eventually a value of the chemical potential $\mu^*$ is
reached for which the two valence states are degenerate, and at
which value the chemical potential is then pinned.\cite{Dzero_2005}
The resulting charge Kondo effect, if present, clearly requires that
hybridization between the impurity states and the host material is
relatively weak.  The semiconducting nature of the host PbTe would
naturally provide an environment in which the local density of
states at the impurity sites is rather small.

Now, we discuss the origin of the enhanced electronic contribution
to the heat capacity seen in Figure 1. The density of states at the
Fermi energy, $N(0)$, can be estimated from the linear term $\gamma$
in the heat capacity.  The resulting values of $N(0)$ are shown on
the right axis of the inset to Figure 1. Part of the observed
enhancement can be attributed to band filling effects, since the
Hall number continues to rise slowly with $x$ even for $x
> x_c$.\cite{paper3} However, as has been discussed
elsewhere,\cite{Chernik_1981b} the observed heat capacity is larger
than expected from the band structure of PbTe alone (dashed line in
the inset to Fig. 1), implying the presence of additional states
associated with the Tl impurities. Within the charge Kondo model,
the additional contribution would arise from the Abrikosov-Suhl
resonance that develops at $E_F$ for temperatures below $T_K$. If
this is the case, then in principle we can estimate the
concentration of Kondo impurities using the crude approximation
$\gamma T_K \sim R$ln2 per mole of impurities. Unfortunately,
uncertainty in the band parameters describing the $\Sigma$
band$^{21}$ means that it is difficult to confidently extract the
magnitude of the additional contribution to the heat capacity over
and above the band-filling effect. Nevertheless, we can make a rough
estimate to at least put limits on the applicability of this model.
Using the measured Hall coefficient,\cite{paper3} published band
parameters,$^{21}$ and the assumption that the band offset does not
change with Tl doping, the band contribution to $\gamma$ can be
estimated as shown by the dashed line in the inset to Fig. 1.  For
$x=1.4\%$, the observed $\gamma$ is 0.46 mJ$\,$mol$^{-1} \,$K$^{-2}$
larger than the expected band contribution.  If this enhancement is
due to Kondo physics, then for $T_K \sim 6$ K (the value estimated
in Ref.~\onlinecite{Matsushita_2005} for $x=0.3\%$), the
concentration of Kondo impurities is approximately $7\times10^{18}$
cm$^{-3}$. In comparison, $x$ = 1.4$\%$ corresponds to a Tl
concentration of $2\times10^{20}$ cm$^{-3}$. Hence, if a charge
Kondo description is appropriate for this material, and if the Kondo
temperature is $\sim 6$ K, then only a small fraction ($\sim3\%$) of
the Tl impurities are contributing to this effect. Within the charge
Kondo model outlined above, this would imply that the Tl impurities
in PbTe must be characterized by a range of $\mu^*$ values, such
that only the subset of impurities for which $\mu = \mu^*$ have
degenerate valence states.

There are other observations that appear to support this tentative
conclusion.  As we had previously noted,\cite{Matsushita_2005} the
magnitude of the resistivity anomaly is also less than what would be
expected if all of the Tl impurities were contributing to the Kondo
effect. Data for lower Tl concentrations, for which a reasonable fit
of the low-temperature data can be made over an extended temperature
range, indicated that approximately 1$\%$ of the Tl impurities
contribute to the Kondo behavior.\cite{Matsushita_2005}  This is in
broad agreement with the value deduced above from the heat capacity
enhancement. In addition, the Hall number is observed to continue to
rise for $x > x_c$,\cite{paper3} implying that the chemical
potential is not pinned at one precise value, but rather is slowed
in its progress as $x$ increases, also consistent with a
distribution of $\mu^*$ values.

Invoking Kondo physics of course implies a temperature dependence to
the enhancement of $\gamma$ for temperatures above $T_K$.  Our
measurements (Fig. 1) show that the enhancement to $\gamma$ is
temperature independent for temperatures between 0.3 K and 1 K.
However, uncertainty in this difference grows rapidly at higher
temperatures due to the increasingly large phonon contribution to
the heat capacity.  As a result, we cannot unambiguously extract the
temperature dependence of any heat capacity enhancement beyond the
estimated Kondo temperature of 6 K.

Within a BCS scenario, $T_c$ varies exponentially with $-1/N(0)V$,
where $V$ is the pairing interaction. Figure 6 shows ln($T_c$)
versus $1/\gamma$ for samples with $x > x_c$. For samples with $T_c
> 0.5$ K, both parameters were extracted from the same physical
crystal. However, for samples with a lower critical temperature,
$T_c$ was determined from resistivity measurements on different
crystals from the same growth batch, introducing additional errors
due to uncertainty in the Tl concentration. As can be seen,
ln($T_c$) scales approximately linearly with $1/\gamma$ within the
uncertainty of the measurements. For a constant $V$, this would
imply that the observed trend in $T_c$ with $x$
(Fig.~\ref{fig:Tcvx}) is due to the increasing density of states
(inset to Fig. 1). However, the situation is less clear if the
charge Kondo picture is applicable, in which case $V$ depends
strongly on the Tl concentration,\cite{Dzero_2005} and the
enhancement in $N(0)$ derives from Kondo physics. In the case of a
superconductor with magnetic impurities, although $N(0)$ is enhanced
by this effect, the superconductivity is nevertheless suppressed for
$T \sim T_K$ due to the pair-breaking effect associated with the
rapid fluctuations in the magnetic moment.\cite{Maple_1972,
Muller-Hartmann_1971} In the case of the charge Kondo model, the
situation is slightly more complex because the impurities now
provide both a local pairing mechanism as well as a pair-breaking
effect close to $T_K$. Consequently, the range of temperatures over
which it is anticipated that $T_c$ will be suppressed is predicted
to be much lower than $T_K$,\cite{Dzero_2005} in contrast to the
case of magnetic impurities. Hence, for the case $T_c \sim T_K$, the
superconductivity can in principle benefit from the enhancement in
$N(0)$ due to the charge Kondo effect in a way that it cannot for
magnetic impurities. The observed trend shown in Fig. 6 may reflect
this effect, but it is difficult to obtain quantitative estimates of
the relative contributions to $T_c$ from the enhancement in $N(0)$
and the pairing interaction itself in this crossover regime of $T_c
\sim T_K$.\cite{Dzero_2005}

\begin{figure}
\includegraphics{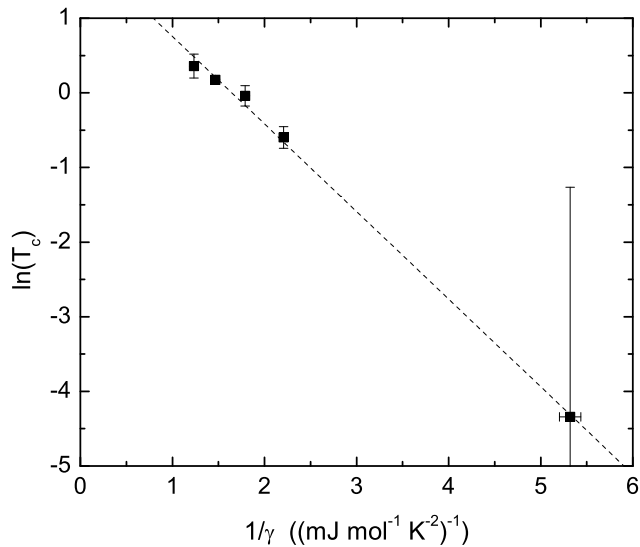}
\caption{\label{fig:lnTc}Plot of ln($T_c$) vs. $1/\gamma$. Dashed
line is a guide for the eye.}
\end{figure}

In the charge Kondo model, if $T_c$ is large compared to $T_K$, then
the pseudospin moment is unscreened at $T_c$, in which case the
superconductivity is born from preformed pairs. In this limit, one
would anticipate a much smaller anomaly in the heat capacity $\Delta
C/\gamma T_c$ than the BCS result of 1.43. As noted in
Section~\ref{sec:iii}, this is clearly not the case for the highest
Tl concentrations, consistent with our previous observation that
$T_c \sim T_K$ for this material.\cite{Matsushita_2005} However, it
is difficult to understand the observed $x$ dependence of $\Delta
C/\gamma T_c$ within this same picture. Since $T_c$ decreases with
decreasing $x$ (Fig. 3), one would expect the superconductivity to
become more BCS-like at lower Tl concentrations. Instead, we find
that $\Delta C/\gamma T_c$ becomes substantially smaller as $x$ is
reduced (inset to Figure 2). Experiments are in progress to measure
the heat capacity of samples with yet smaller Tl concentrations to
even lower temperatures to see whether this trend continues.

Could the superconductivity in Tl-doped PbTe have its origin in more
mundane physics after all?  While the data presented here enable us
to characterize this material as a BCS superconductor, they do not
allow us to distinguish between different pairing mechanisms. As we
have previously argued,\cite{Matsushita_2005} many aspects of the
observed thermodynamic and transport properties are suggestive of
charge Kondo physics.  Moreover, the uniqueness of the Tl
impurities, being the only dopant to cause superconductivity in this
material, cannot be ignored. Nevertheless, in the absence of
experiments directly probing the Tl valence (which are currently in
progress), we cannot rule out less exciting possibilities, including
the formation of a narrow impurity band with a relatively large
density of states. In such a case, the pairing interaction would
most likely be phonon mediated, though the substantial residual
resistance might argue that strong Coulomb scattering also plays a
role. The observed low-temperature resistivity anomaly would then
presumably have its origin in some form of weak localization, though
the temperature and field dependence of this feature appear to argue
against such a scenario.\cite{Matsushita_2005}

\section{Conclusions}
In summary, we have shown that Tl-doped PbTe is a Type II, BCS
superconductor in the dirty limit. None of these observations is in
disagreement with the charge Kondo model previously described,
though they do put some limitations on its applicability.
Specifically, the relatively small enhancement of the electronic
contribution to the heat capacity implies that if a charge Kondo
description is appropriate then only a small fraction of the Tl
impurities can be participating in the Kondo physics. Within the
model described in Ref.~\onlinecite{Dzero_2005}, this can be
understood in terms of a distribution of $\mu$* values, such that
only a subset of the Tl impurities have both valence states exactly
degenerate for a particular value of the chemical potential within
this range, though this has yet to be experimentally verified.

\begin{acknowledgments}
We gratefully thank J.~Schmalian, M.~Dzero, M.~R. Beasley, and
B.~Moyzhes for numerous helpful discussions.  We also acknowledge
Robert E. Jones for technical assistance with EMPA measurements and
A.~T. Sommer for help with Hall effect measurements. This work is
supported by the DOE, Office of Basic Energy Sciences, under
contract number DE-AC02-76SF00515. I.~R.~F was also supported by the
Alfred P.~Sloan and Terman Foundations.
\end{acknowledgments}


\bibliography{Matsushita2_condmat}

\end{document}